# Anomalous Plasticity of Body-Centered-Cubic Crystals with Non-Schmid Effect

Hansohl Cho[†], Curt A. Bronkhorst[‡], Hashem M. Mourad, Jason R. Mayeur and D. J. Luscher

Theoretical Division, Los Alamos National Laboratory, Los Alamos, NM 87545, US

Email: [†]hansohl.cho@gmail.com; [‡]cabronk@lanl.gov

**Abstract**

Plastic deformations in body-centered-cubic (BCC) crystals have been of critical importance in diverse engineering and manufacturing contexts across length scales. Numerous experiments and atomistic simulations on BCC crystals reveal that classical crystal plasticity models with the Schmid law are not adequate to account for abnormal plastic deformations often found in these crystals. In this paper, we address a continuum mechanical treatment of anomalous plasticity in BCC crystals exhibiting non-Schmid effects, inspired from atomistic simulations recently reported. Specifically, anomalous features of plastic flows are addressed in conjunction with a single crystal constitutive model involving two non-Schmid projection tensors widely accepted for representing non-glide components of an applied stress tensor. Further, modeling results on a representative BCC single crystal (tantalum) are presented and compared to experimental data at a range of low temperatures to provide physical insight into deformation mechanisms in these crystals with non-Schmid effects.

## 1. Introduction

Crystal plasticity has been a focal point of research since it is of critical importance in processing and manufacturing of crystalline materials at a wide range of deformation conditions. Over the past several decades, numerous crystal plasticity theories have been proposed to model the deformations of single- and polycrystals, based on microscopic mechanisms such as dislocation slip, twinning and phase transformation (Asaro, 1983a; Asaro and Needleman, 1985; Hansen et al., 2013; Kalidindi et al., 1992; Staroselsky and Anand, 2003; Thamburaja and Anand, 2001). Moreover, computational procedures for crystal plasticity models have been widely investigated for use in numerical solvers for boundary value problems (Anand and Kothari, 1996; Cuitino and Ortiz, 1993; Kalidindi et al., 1992; Miehe, 1996).

The microscopic "slip" of atomic configurations has been central to many continuum crystal plasticity theories. The classical crystal plasticity based on a shear stress projected via a Schmid tensor ($\mathbf{S}_0 = \mathbf{m}_0 \otimes \mathbf{n}_0$, $\mathbf{m}_0$: slip direction, and $\mathbf{n}_0$: slip plane normal) has enabled to capture salient features of slip-based plastic deformations especially in close-packed crystals. In the Schmid law, the critical resolved shear stress for onset of inelastic slips is assumed to be dependent only upon a shear stress on a slip plane in a slip direction. However, in many crystals, the Schmid law was found to break down with spreading of dislocation cores onto several non-parallel planes, resulting in the stress-state dependence of the Peierls barrier. Such non-planar cores of dislocations are mainly attributed to complex crystallographic aspects, mixed nature of metallic and covalent bonding and the interplay between them especially in the BCC transition metals of group V and VI (Duesbery et al., 1973; Duesbery and Vitek, 1998b; Nabarro and Duesbery, 2002; Vitek, 1974; Vitek et al., 1970). The breakdown of the Schmid law gives rise to abnormal plasticity such as tension-compression asymmetry and orientation-dependent critical stress, as is well reviewed in the literature (Duesbery and Vitek, 1998a; Ito and Vitek, 2001; Vitek, 2004). Moreover, localization and bifurcation of shear bands, often resulting in ductile failures, have been widely studied in experiments and numerical simulations with non-Schmid effects (Asaro and Rice, 1977; Dao and Asaro, 1993; Peirce et al., 1983; Racherla and Bassani, 2006). The geometry of screw dislocation cores in BCC crystals is the most widely investigated example, in which non-Schmid effects are invoked by non-glide stresses. The non-glide stresses and their connections to the altered core of screw dislocations have been investigated by *ab-initio* (Dezerald et al., 2014; Frederiksen and Jacobsen, 2003; Ismail-Beigi and Arias, 2000; Woodward and Rao, 2002) and atomistic simulations (Gröger et al., 2008a; Li et al., 2004; Mrovec et al., 2004; Ravelo et al., 2013; Vitek, 2004; Vitek et al., 2004a; Xu and Moriarty, 1996). Specifically, diverse interatomic force fields such as embedded atomic method (EAM), bond-order potentials, MGPT and their variations were employed to address the geometric details of a non-planar core of $1/2\langle 111 \rangle$ screw dislocations. Non-Schmid effects have also been experimentally evidenced in many BCC transition metals (Byron, 1968; Carroll et al., 2013; Ferriss et al., 1962; Irwin et al., 1974; Kim et al., 2010; Sherwood et al., 1967), in which orientation-dependent tension-compression asymmetries and anomalous yield loci were reported across length-scales. Moreover, such unusual plastic deformations have been reported in some FCC and HCP metals (Dao and Asaro, 1993; Lim et al., 2013).

In this paper, we present a constitutive modeling framework that accounts for non-Schmid effects by employing two representative projection tensors introduced by Vitek, Bassani and co-workers (Gröger et al., 2008a; Qin and Bassani, 1992a, b; Vitek et al., 2004a; Vitek et al., 2004b) and Asaro, Rice and co-workers (Asaro, 1983b; Asaro and Rice, 1977; Dao and Asaro, 1993). Atomistic simulations of single crystals constituted a physical foundation to the former with a secondary projection plane ($\mathbf{n}'_0$) for non-glide stresses. The latter was more empirically established, using additional dyadic combinations of a slip plane and a slip direction which lead to five non-glide stresses. With both formulations, comprehensive modeling studies for crystal plasticity are performed, by which anomalous plastic features in these crystals are elucidated. First, a comparative parametric study for the two approaches shows how each of the non-glide stresses related to the two tensors influences the abnormal yield under various shear stresses. Second, a single crystal constitutive model is specialized to simulate the stress-strain behavior of a BCC crystal, by which unique features of tension-compression asymmetry are addressed in different crystallographic orientations. Then, the constitutive model is further addressed to capture temperature-dependent non-Schmid effects. Finally, stress-strain curves of a representative tantalum are modeled at a range of low temperatures; and compared to experimental data, revealing the ability of the constitutive model to capture the main features of plastic flows with non-Schmid effects.

## 2. Non-Schmid Effect in BCC Crystals : Implications from Atomistic Simulations

Atomistic simulations have been widely used to address plastic deformation in many BCC metals, revealing that the non-planar core of screw dislocations results in non-Schmid effects. Various spreading patterns of the dislocation cores were observed in the atomistic simulations, highly dependent on the interatomic potentials. Vitek, Duesbery and coworkers found that the energy landscape (γ-energy surface) of a generalized stacking fault for the major slip planes plays a decisive role in determining the core patterns (Duesbery and Vitek, 1998a; Ito and Vitek, 2001; Vitek, 2004). Specifically, two distinct spreading patterns were extensively observed in their atomistic simulations: (1) degenerate three-fold symmetric and (2) non-degenerate six-fold symmetric patterns in the zone of a $1/2\langle 111\rangle$ screw dislocation. Notably, the overall behavior of $1/2\langle 111\rangle$ screw dislocations under the effect of an applied stress tensor was similar for both patterns. The original symmetry associated with the $[10\bar{1}]$ diad is eliminated and, consequently, the distinction between the two spreading patterns may vanish, prior to dislocation glide. Moreover, the non-glide stress components of an applied stress tensor change the critical shear stress by modifying the core spreading pattern, but they never produce the Peach-Koehler force; i.e., though the non-glide stresses cannot cause any dislocation glide, they give rise to a significant change in the critical shear stress for onset of dislocation slip.

**Figure 1** summarizes recent atomistic simulation results (open and closed symbols) of an isolated screw dislocation in single crystal tungsten (Gröger et al., 2008a), molybdenum (Vitek et al., 2004a) and α-iron (Lim et al., 2015) under shear stress at 0 K. Provided that the Schmid law holds, the critical shear stress should be symmetric about an angle ($\varphi$)[a] between the slip plane and the maximum resolved shear stress plane (MRSSP), as shown in the atomistic simulation of tungsten in **Figure 1b**.[b] Meanwhile, it was highly asymmetric for molybdenum and α-iron as shown in **Figure 1c** and **d**, which was attributed to the non-glide stress components of an applied stress tensor.

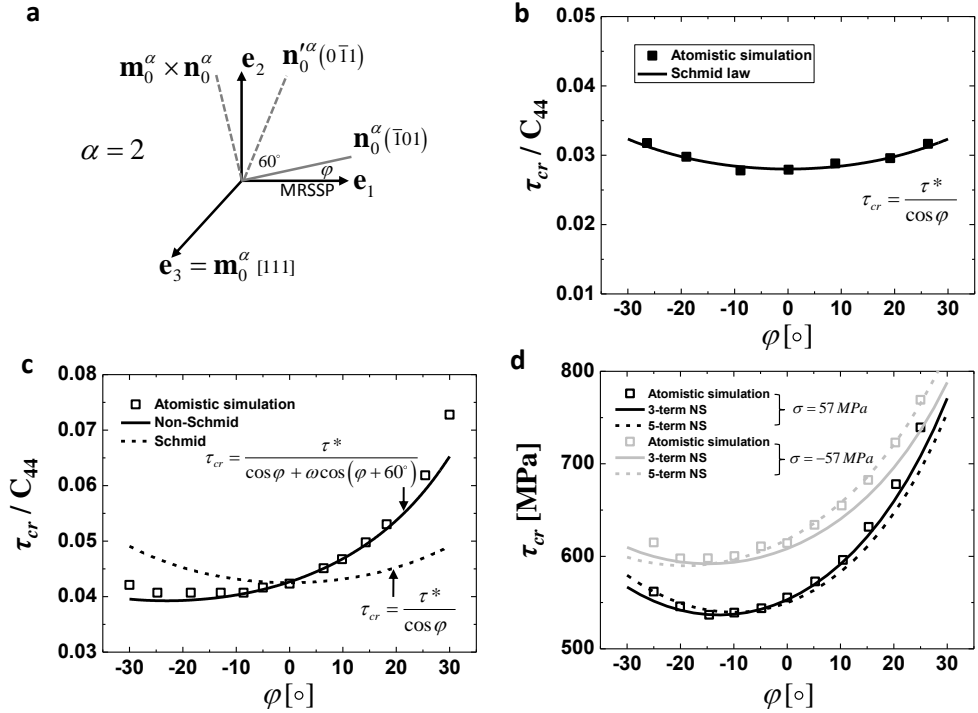

**Figure 1** Orientation-dependent critical shear stress ($\tau_{cr}$) by non-Schmid effects. (a) Coordinate system for a representative slip system of $\alpha = 2$ ($\mathbf{e}_2$ is normal to MRSSP), (b) $\tau_{cr}$ as a function of $\varphi$ (positive: clockwise from $\mathbf{n}_0^\alpha$) in tungsten, (c) $\tau_{cr}$ in molybdenum, (d) $\tau_{cr}$ in α-iron

### 2.1. Non-Schmid Projection Tensors for Non-Glide Stresses

In the Schmid law, a resolved shear stress (RSS) that drives inelastic slip is approximated by $\tau^\alpha \approx \mathbf{T}^* : \mathbf{S}_0^\alpha$. Here, $\mathbf{T}^*$ is the second Piola-Kirchhoff stress in intermediate space (also referred to as lattice space), $\mathbf{S}_0^\alpha = \mathbf{m}_0^\alpha \otimes \mathbf{n}_0^\alpha$ is the Schmid tensor that defines $\alpha^{th}$ slip system, $\otimes$ denotes a tensor product of two vector, : denotes a scalar product of two tensors, $\mathbf{m}_0^\alpha$ is the slip direction, and $\mathbf{n}_0^\alpha$ is the slip plane normal with $\mathbf{m}_0^\alpha \cdot \mathbf{n}_0^\alpha = 0$ and $|\mathbf{m}_0^\alpha| = |\mathbf{n}_0^\alpha| = 1$ in the intermediate space. For non-Schmid effects, the resolved shear stress should be modified to have non-glide stress contributions, as follows,

$$\tilde{\tau}^\alpha = \mathbf{T}^* : \left(\mathbf{S}_0^\alpha + \tilde{\mathbf{S}}_0^\alpha\right), \tag{1}$$

where $\tilde{\mathbf{S}}_0^\alpha$ is the secondary projection tensor defined in each of the slip systems.

### 2.1.1. Three-term Formulation

Vitek, Bassani and coworkers introduced a secondary projection plane to modify the resolved shear stress especially for slip systems of $\{110\}\langle 111\rangle$, guided from atomistic simulations (Gröger et al., 2008a; Vitek et al., 2004a; Vitek et al., 2004b). It has been widely used for modeling plastic flows in BCC crystals, giving a physical insight into the geometric nature of the altered dislocation cores and non-glide stresses (Alleman et al., 2014; Koester et al., 2012; Patra et al., 2014). With this plane, we may have another projection tensor for a non-glide stress, defined as $\tilde{\mathbf{S}}_0^\alpha = \omega \mathbf{m}_0^\alpha \otimes \mathbf{n}_0^{\prime\alpha}$. Here, $\omega$ is the weighting factor and $\mathbf{n}_0^{\prime\alpha}$ is a unit vector normal to the secondary projection plane, whose angle with the original slip plane normal, $\mathbf{n}_0^\alpha$, is $60°$. The modified resolved shear stress with $\mathbf{S}_0^\alpha + \tilde{\mathbf{S}}_0^\alpha$ captured successfully the asymmetric critical shear stress ($\tau_{cr}$) about the angle ($\varphi$) in molybdenum as shown in **Figure 1c**. Here, the critical shear stress on MRSSP was computed from the atomistic simulations (open symbols) under $\mathbf{T}|_{\{\mathbf{e}_i\}} = 2\tau \cdot \text{sym}(\mathbf{e}_2 \otimes \mathbf{e}_3)$. Also, the critical shear stress (solid line) is simply expressed by $\tau_{cr} = \dfrac{\tau^*}{\cos\varphi + \omega\cos(\varphi + 60°)}$ with fitting parameters of $\tau^*/C_{44} = 0.0565$ ($\tau^*/C_{44} = 0.0425$ for the Schmid law: dashed line) and $\omega = 0.65$ ($\omega > 0$ for twin/anti-twin asymmetry), where $\tau^*$ is the shear yield stress in this particular slip system.

More recently, the non-Schmid projection tensor has been furthered to involve additional non-glide stresses (Gröger et al., 2008a; Gröger et al., 2008b), as follows,

$$\tilde{\mathbf{S}}_0^\alpha = \sum_{i=1}^{3} \omega_i \tilde{\mathbf{S}}_0^{i,\alpha} \text{ with } \tilde{\mathbf{S}}_0^{1,\alpha} = \mathbf{m}_0^\alpha \otimes \mathbf{n}_0'^\alpha, \ \tilde{\mathbf{S}}_0^{2,\alpha} = \left(\mathbf{n}_0^\alpha \times \mathbf{m}_0^\alpha\right) \otimes \mathbf{n}_0^\alpha, \ \tilde{\mathbf{S}}_0^{3,\alpha} = \left(\mathbf{n}_0'^\alpha \times \mathbf{m}_0^\alpha\right) \otimes \mathbf{n}_0'^\alpha, \quad (2)$$

where $\tilde{\mathbf{S}}_0^{1,\alpha}$ represents the projection tensor for a non-glide stress "parallel" to the slip direction while $\tilde{\mathbf{S}}_0^{2,\alpha}$ and $\tilde{\mathbf{S}}_0^{3,\alpha}$ represent the projection tensors for non-glide stresses "perpendicular" to the slip direction. The non-glide stresses via these projection tensors account for asymmetry in the critical shear stress of a BCC crystal under more complex loading conditions, as shown for α-iron in **Figure 1d**. The critical shear stress ($\tau_{cr}$) on MRSSP was computed from the atomistic simulations (open symbols) under a combined loading condition, $\mathbf{T}|_{\{\mathbf{e}_i\}} = -\sigma \mathbf{e}_1 \otimes \mathbf{e}_1 + \sigma \mathbf{e}_2 \otimes \mathbf{e}_2 + 2\tau \cdot \text{sym}(\mathbf{e}_2 \otimes \mathbf{e}_3)$, with $\sigma$ and $\tau$ perpendicular and parallel to the slip direction. Under this loading condition, the critical shear stress decreased as $\sigma$ increased; i.e. the Peierls stress decreased due to the altered dislocation core with increased $\sigma$, which is perpendicular to the slip direction. $\tau_{cr}$ (solid line) on MRSSP is simply expressed by (see **Appendix A**),

$$\tau_{cr} = \frac{\tau^* - \sigma\left(\omega_2 \sin 2\varphi + \omega_3 \sin 2\left(\varphi + 60°\right)\right)}{\cos\varphi + \omega_1 \cos\left(\varphi + 60°\right)}. \tag{3}$$

where $\tau^* = 675\, MPa$, $\omega_1 = 0.325$, $\omega_2 = 0.2$ and $\omega_3 = 0.65$.

Further, **Figure 2a – c** shows effects of the non-Schmid projection tensors and the relevant non-glide stresses on the critical shear stress in Equation (3) under the combined loading condition. **Figure 2a** shows a contour for a normalized critical stress on MRSSP with varying $\omega_1 \in [0,1]$ and $\varphi \in [-30°, 30°]$. As expected, asymmetry in $\tau_{cr}$ became stronger with increased contribution from this non-glide stress. Effects of non-glide stresses perpendicular to the slip direction are examined in **Figure 2b** and **Figure 2c** with varying $\sigma/\tau^*$. As $\sigma$ increased, a change in $\tau_{cr}$ due to $\omega_{2,3}$ increased significantly in both $\varphi=-15°$ and $\varphi=15°$. Moreover, for $\varphi>0°$, as $\sigma$ increased, $\tau_{cr}$ was found to decrease (more likely to yield). Interestingly, for $\varphi<0°$, $\tau_{cr}$ decreased as the non-glide stress via $\tilde{\mathbf{S}}_0^{3,\alpha}$ increased. However, it increased with an increased non-glide stress via $\tilde{\mathbf{S}}_0^{2,\alpha}$, thereby revealing another significant asymmetry in $\tau_{cr}$ due to these non-glide stresses perpendicular to the slip direction.

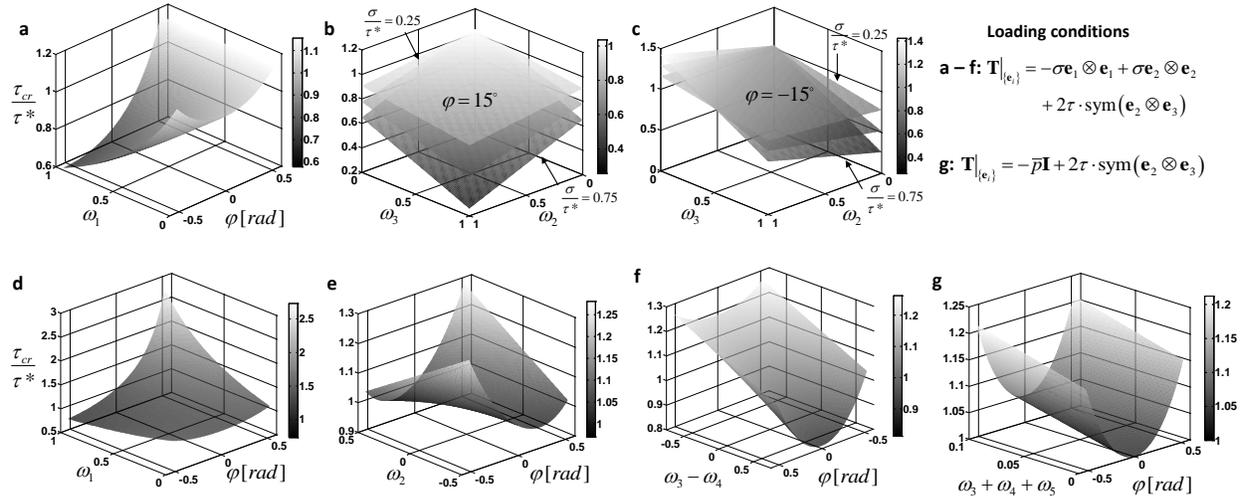

**Figure 2** Effect of non-Schmid stresses on $\tau_{cr}$. (a) $\tau_{cr}$ vs. $\varphi$ and $\omega_1$ ($\omega_{2,3}=0$); (b) $\tau_{cr}$ vs. $\omega_2$ and $\omega_3$ with $\varphi=15°$ ($\omega_1=0$); and (c) $\tau_{cr}$ vs. $\omega_2$ and $\omega_3$ with $\varphi=-15°$ ($\omega_1=0$) in 3-term formulation, (d) $\tau_{cr}$ vs. $\varphi$ and $\omega_1$ ($\omega_{i\neq 1}=0$); (e) $\tau_{cr}$ vs. $\varphi$ and $\omega_2$ ($\omega_{i\neq 2}=0$ and $\sigma/\tau^*=0.25$); (f) $\tau_{cr}$ vs. $\varphi$ and $\omega_3-\omega_4$ ($\omega_{i=1,2}=0$ and $\sigma/\tau^*=0.25$) and (g) $\tau_{cr}$ vs. $\varphi$ and $\omega_3+\omega_4+\omega_5$ ($\omega_{i=1,2}=0$ and $\bar{p}/\tau^*=0.5$) in 5-term formulation.

### 2.1.2. Five-term Formulation

The non-Schmid projection tensor was also specialized by Asaro, Rice and coworkers, using dyadic combinations of a slip direction and a slip plane normal (Asaro, 1983b; Asaro and Rice, 1977; Dao and Asaro, 1993), as follows,

$$\tilde{\mathbf{S}}_0^\alpha = \sum_{i=1}^{5} \omega_i \tilde{\mathbf{S}}_0^{i,\alpha} \text{ with } \tilde{\mathbf{S}}_0^{1,\alpha} = \mathbf{l}_0^\alpha \otimes \mathbf{m}_0^\alpha, \tilde{\mathbf{S}}_0^{2,\alpha} = \mathbf{l}_0^\alpha \otimes \mathbf{n}_0^\alpha, \tilde{\mathbf{S}}_0^{3,\alpha} = \mathbf{n}_0^\alpha \otimes \mathbf{n}_0^\alpha, \tilde{\mathbf{S}}_0^{4,\alpha} = \mathbf{l}_0^\alpha \otimes \mathbf{l}_0^\alpha, \tilde{\mathbf{S}}_0^{5,\alpha} = \mathbf{m}_0^\alpha \otimes \mathbf{m}_0^\alpha, \quad (4)$$

where $\mathbf{l}_0^\alpha = \mathbf{m}_0^\alpha \times \mathbf{n}_0^\alpha$. In this formulation, the non-glide "shear" stresses via $\tilde{\mathbf{S}}_0^{1,\alpha}$ and $\tilde{\mathbf{S}}_0^{2,\alpha}$ are only perpendicular to the slip direction; i.e., there is no projection tensor for twin/anti-twin asymmetry (e.g. $\mathbf{m}_0^\alpha \otimes \mathbf{n}_0'^\alpha$ in the three-term formulation). However, the five-term formulation has projection tensors ($\tilde{\mathbf{S}}_0^{3,\alpha}$, $\tilde{\mathbf{S}}_0^{4,\alpha}$ and $\tilde{\mathbf{S}}_0^{5,\alpha}$) for the effects of "normal" stresses. Their sum is usually taken to be "deviatoric"; i.e., $\omega_3 + \omega_4 + \omega_5 = 0$. Meanwhile, a non-zero $\omega_3 + \omega_4 + \omega_5$ is a pressure sensitivity coefficient.

This five-term formulation is empirical with no rigorous physical background for each of the additional projection tensors. However, it has enabled to model non-Schmid effects in BCC crystals at a continuum level without any additional projection "plane" (e.g., $\mathbf{n}_0'^\alpha$ in the three-term formulation) for the modified RSS (Dao and Asaro, 1993; Knezevic et al., 2014; Lim et al., 2013; Savage et al., 2017). Moreover, as shown for α-iron in **Figure 1d**, $\tau_{cr}$ computed from the atomistic simulations (open symbol in **Figure 1d**) under $\mathbf{T}|_{\{\mathbf{e}_i\}} = -\sigma \mathbf{e}_1 \otimes \mathbf{e}_1 + \sigma \mathbf{e}_2 \otimes \mathbf{e}_2 + 2\tau \cdot \text{sym}(\mathbf{e}_2 \otimes \mathbf{e}_3)$, was nicely captured by the five-term formulation, where $\tau_{cr}$ (dashed line in **Figure 1d**) on MRSSP is simply expressed by,

$$\tau_{cr} = \frac{\tau^* + \sigma(\omega_2 \sin 2\varphi - (\omega_3 - \omega_4)\cos 2\varphi)}{\cos\varphi - \omega_1 \sin\varphi}, \quad (5)$$

where $\tau^* = 584\, MPa$, $\omega_1 = 0.25$, $\omega_2 = -0.2$ and $\omega_3 - \omega_4 = 0.55$. Here, it should be noted that the sign of $\omega_2$ was chosen to be opposite to that in the three-term formulation since $\tilde{\mathbf{S}}_0^{2,\alpha}|_{3-term} = -\tilde{\mathbf{S}}_0^{2,\alpha}|_{5-term}$ (see solid and dashed lines in **Figure 1d** for comparison of the three-term and five-term formulations).

Non-glide stresses via the five-term formulation are further addressed under two different loading conditions. First, $\tau_{cr}$ is examined under $\mathbf{T}|_{\{\mathbf{e}_i\}} = -\sigma \mathbf{e}_1 \otimes \mathbf{e}_1 + \sigma \mathbf{e}_2 \otimes \mathbf{e}_2 + 2\tau \cdot \text{sym}(\mathbf{e}_2 \otimes \mathbf{e}_3)$ in **Figure 2d**

– **f** with a range of weighting factors. **Figure 2d** and **Figure 2e** show contours of a normalized $\tau_{cr}$ with varying $\omega_1$ (or $\omega_2$) and $\varphi$. As expected, asymmetry in $\tau_{cr}$ became stronger as the non-glide stresses via $\tilde{\mathbf{S}}_0^{1,\alpha}$ and $\tilde{\mathbf{S}}_0^{2,\alpha}$ increased. Moreover, anti-symmetry in $\tau_{cr}$ about the sign of the weighting factor was apparent, as shown in **Figure 2e**. Notably, the non-Schmid effect vanished at $\varphi = 0°$ for these non-glide stresses, which was contrary to that observed in the three-term formulation (see **Figure 2a**).[c] Moreover, $\tau_{cr}$ by varying $\omega_3 - \omega_4$ and $\varphi$ is presented in **Figure 2f**. Normal stress difference via $\tilde{\mathbf{S}}_0^{3,\alpha}$ and $\tilde{\mathbf{S}}_0^{4,\alpha}$ influenced remarkably $\tau_{cr}$, which is symmetric about $\varphi$; i.e., when $\omega_3 = \omega_4$, the applied stress ($\sigma$) parallel to the slip direction does not have any effect on $\tau_{cr}$.

$\tau_{cr}$ is then examined under a "non-deviatoric" loading condition, $\mathbf{T}|_{\{\mathbf{e}_i\}} = -\bar{p}\mathbf{I} + 2\tau \cdot \text{sym}(\mathbf{e}_2 \otimes \mathbf{e}_3)$ with the positive mean normal pressure, $\bar{p} = -tr\mathbf{T}/3$. Here, $\tau_{cr}$ is simply expressed by

$$\tau_{cr} = \frac{\tau^* + \bar{p}(\omega_3 + \omega_4 + \omega_5)}{\cos\varphi - \omega_1 \sin\varphi}.$$ **Figure 2g** shows a contour of $\tau_{cr}$ with varying pressure sensitivity ($\omega_3 + \omega_4 + \omega_5$) and $\varphi$. Obviously, $\tau_{cr}$ under finite pressure was symmetric about $\varphi$. Moreover, it increased monotonically with increased pressure sensitivity; i.e., the "normal" stresses via $\tilde{\mathbf{S}}_0^{3,\alpha}$, $\tilde{\mathbf{S}}_0^{4,\alpha}$ and $\tilde{\mathbf{S}}_0^{5,\alpha}$ enable to represent a pressure-sensitive plastic flow.

### 2.2. Tension-compression Asymmetry: Analysis for a Single Slip System

The modified resolved shear stress influences significantly tension-compression asymmetry often found in BCC crystals under uniaxial loading. Here, we show how each of the non-Schmid stresses influences the tension-compression asymmetry for a single slip system ($\alpha$) and its conjugate pair ($\alpha + 12$) for both three- and five-term formulations. For brevity, the non-Schmid stress for each projection tensor is defined as, $\tilde{\tau}_{NS}^{i,\alpha} = \mathbf{T} : \tilde{\mathbf{S}}_0^{i,\alpha}$, where $\mathbf{T}$ is an uniaxial compression or tension. Moreover, if $\alpha$ is active, $\alpha+12$ is naturally inactive; or *vice versa*. (see **Appendix C** for detail)

In the five term formulation, from $\tilde{\mathbf{S}}_0^{2,\alpha} = (\mathbf{m}_0^\alpha \times \mathbf{n}_0^\alpha) \otimes \mathbf{n}_0^\alpha$ and $\tilde{\mathbf{S}}_0^{2,\alpha+12} = (\mathbf{m}_0^{\alpha+12} \times \mathbf{n}_0^\alpha) \otimes \mathbf{n}_0^\alpha = -\tilde{\mathbf{S}}_0^{2,\alpha}$, we have $\tilde{\tau}_{NS}^{2,\alpha}\big|_{tens} = \tilde{\tau}_{NS}^{2,\alpha+12}\big|_{comp}$; i.e., this non-Schmid stress does not contribute to any asymmetry, similar to the standard RSS. However, the yield stress in both tension and compression may increase or decrease simultaneously due to $\tilde{\tau}_{NS}^{2,\alpha}$ (or $\tilde{\tau}_{NS}^{2,\alpha+12}$), depending on the weighting factor. However, from $\tilde{\mathbf{S}}_0^{1,\alpha} = (\mathbf{m}_0^\alpha \times \mathbf{n}_0^\alpha) \otimes \mathbf{m}_0^\alpha$ and $\tilde{\mathbf{S}}_0^{1,\alpha+12} = (\mathbf{m}_0^{\alpha+12} \times \mathbf{n}_0^\alpha) \otimes \mathbf{m}_0^{\alpha+12} = \tilde{\mathbf{S}}_0^{1,\alpha}$, we have $\tilde{\tau}_{NS}^{1,\alpha}\big|_{tens} = -\tilde{\tau}_{NS}^{1,\alpha+12}\big|_{comp}$; i.e., this non-Schmid stress contributes significantly to asymmetric yield in tension and compression. A difference between yield stresses with and without this non-Schmid stress is naturally identical in both compression and tension. Additionally, the normal stresses for $\tilde{\mathbf{S}}_0^{3,\alpha}$, $\tilde{\mathbf{S}}_0^{4,\alpha}$ and $\tilde{\mathbf{S}}_0^{5,\alpha}$ contribute to asymmetry since $\tilde{\tau}_{NS}^{(i=3,4,5),\alpha}\big|_{tens} = -\tilde{\tau}_{NS}^{(i=3,4,5),\alpha+12}\big|_{comp}$ and $\tilde{\mathbf{S}}_0^{(i=3,4,5),\alpha} = \tilde{\mathbf{S}}_0^{(i=3,4,5),\alpha+12}$.

In the three term formulation, since $\tilde{\mathbf{S}}_0^{2,\alpha} = -\tilde{\mathbf{S}}_0^{2,\alpha+12}$, $\tilde{\tau}_{NS}^{2,\alpha}$ does not contribute to any asymmetry but changes the magnitude of the yield stress for both tension and compression, which was the same in the five-term formulation ($\tilde{\mathbf{S}}_0^{2,\alpha}\big|_{3-term} = -\tilde{\mathbf{S}}_0^{2,\alpha}\big|_{5-term}$). For $\tilde{\mathbf{S}}_0^{1,\alpha} = \mathbf{m}_0^\alpha \otimes \mathbf{n}_0^{\prime\alpha}$ and $\tilde{\mathbf{S}}_0^{3,\alpha} = (\mathbf{n}_0^{\prime\alpha} \times \mathbf{m}_0^\alpha) \otimes \mathbf{n}_0^{\prime\alpha}$, the non-Schmid stresses result in asymmetry. However, unlike in the five-term formulation, $\tilde{\tau}_{NS}^{(i=1,3),\alpha}\big|_{tens} \neq -\tilde{\tau}_{NS}^{(i=1,3),\alpha+12}\big|_{comp}$ ($\tilde{\mathbf{S}}_0^{(i=1,3),\alpha} \neq \tilde{\mathbf{S}}_0^{(i=1,3),\alpha+12}$); i.e., a difference between yield stresses with and without these non-Schmid stresses in compression is different from that in tension, enabling to capture more complicated tension-compression asymmetries in the materials. Moreover, in both formulations, the tension-compression asymmetry is naturally captured by the non-Schmid stresses without any pressure sensitivity which leads to asymmetric yield stresses in isotropic plastic solids. The tension-compression asymmetry is further addressed for single crystals with multiple slip systems in **Section 3**.

## 3. Stress-strain Behaviors of BCC Crystal with non-Schmid Effect: Application to Tantalum

In this section, we present a constitutive modeling framework that accounts for non-Schmid effects in BCC single crystals for both three-term and five-term formulations. The single crystal plasticity model is then specialized to simulate the stress-strain behaviors of a representative BCC tantalum that exhibits significant non-Schmid effects along different crystallographic orientations at a range of low temperatures.

### 3.1. Single Crystal Plasticity: a Continuum Model

Here, continuum theories for single crystal plasticity are briefly outlined (Anand, 2004; Gurtin, 2000; Gurtin et al., 2010).

- **Kinematics**

The deformation gradient decomposes multiplicatively,

$$\mathbf{F} \equiv \nabla \boldsymbol{\chi} = \mathbf{F}^e \mathbf{F}^p, \tag{6}$$

where $\nabla(\cdot)$ denotes a gradient in the undeformed reference configuration and $\boldsymbol{\chi}$ is the motion of a material vector; $\mathbf{F}^p$ and $\mathbf{F}^e$ represent the plastic and elastic distortions, respectively. The plastic distortion is assumed to be incompressible in finite deformation elastic-plasticity, i.e., $\det \mathbf{F}^p = 1$.

The spatial velocity gradient describes the rate of deformation, as follows,

$$\mathbf{L} \equiv \operatorname{grad} \mathbf{v} = \dot{\mathbf{F}} \mathbf{F}^{-1}, \tag{7}$$

where "grad" denotes a gradient in the deformed configuration and $\mathbf{v}$ is the spatial velocity field. The velocity gradient decomposes additively into elastic and plastic distortion rate tensors as follows,

$$\mathbf{L} = \mathbf{L}^e + \bar{\mathbf{L}}^p = \mathbf{L}^e + \mathbf{F}^e \mathbf{L}^p \mathbf{F}^{e-1} = \dot{\mathbf{F}}^e \mathbf{F}^{e-1} + \mathbf{F}^e \dot{\mathbf{F}}^p \mathbf{F}^{p-1} \mathbf{F}^{e-1}. \tag{8}$$

The evolution of $\mathbf{F}^p$ is then rearranged to,

$$\dot{\mathbf{F}}^p = \mathbf{L}^p \mathbf{F}^p \text{ with } \mathbf{L}^p = \sum_{\alpha=1}^{N} \dot{\gamma}_p^\alpha \mathbf{S}_0^\alpha, \tag{9}$$

where $\mathbf{S}_0^\alpha = \mathbf{m}_0^\alpha \otimes \mathbf{n}_0^\alpha$ is the Schmid tensor, $\dot{\gamma}_p^\alpha$ is the plastic shear strain rate and $N$ is the number of slip systems. Equivalently, the plastic distortion rate ($\bar{\mathbf{L}}^p$) can be expressed with respect to the deformed configuration ($\bar{\mathbf{L}}^p = \sum_{\alpha=1}^{N} \dot{\gamma}_p^\alpha \mathbf{S}^\alpha$ with $\mathbf{S}^\alpha = \mathbf{m}^\alpha \otimes \mathbf{n}^\alpha$, $\mathbf{m}^\alpha = \mathbf{F}^e \mathbf{m}_0^\alpha$, $\mathbf{n}^\alpha = \mathbf{F}^{e-T} \mathbf{n}_0^\alpha$). For the single crystal plasticity model, we used representative slip systems ($\alpha = 1 \sim 24$) given in **Appendix C**.

- **Constitutive Model**

To describe elastic responses of a single crystal, a quadratic free energy per unit reference volume was employed in the intermediate space as follows,

$$\Phi = \tilde{\Phi}(\mathbf{E}^e) = \frac{1}{2} \mathbf{E}^e : \mathcal{C}^c \mathbf{E}^e, \ \mathbf{T}^* = \frac{\partial \tilde{\Phi}(\mathbf{E}^e)}{\partial \mathbf{E}^e} = \mathcal{C}^c \mathbf{E}^e \tag{10}$$

where $\mathcal{C}^c$ is the fourth order elasticity tensor with cubic symmetry and $\mathbf{E}^e = \frac{1}{2}(\mathbf{C}^e - \mathbf{I})$ is the elastic Green-Lagrangian strain measure ($\mathbf{I}$ is the second order identity tensor) in the intermediate space, where $\mathbf{C}^e = \mathbf{F}^{eT} \mathbf{F}^e$ is the elastic right Cauchy-Green tensor. Moreover, $\mathbf{T}^*$ is the symmetric second Piola-Kirchhoff stress, which is the work-conjugate stress measure to $\mathbf{E}^e$ in the intermediate space. It is also related to the Cauchy stress ($\mathbf{T}$) by $\mathbf{T} = \frac{1}{J} \mathbf{F}^e \mathbf{T}^* \mathbf{F}^{eT}$, where $J = \det \mathbf{F} = \det \mathbf{F}^e$ is the volume change.

Viscoplastic flow is constitutively prescribed by a thermally-activated glide model (Busso, 1990; Kocks, 1976),

$$\dot{\gamma}_p^\alpha = \dot{\gamma}_p^0 \exp\left(-\frac{\Delta G}{k_B \theta}\left\langle 1 - \left\langle \frac{\tau_{eff}^\alpha}{\tilde{s}_l^\alpha}\right\rangle^p\right\rangle^q\right) \text{ for } \tau^\alpha > 0, \text{ otherwise } \dot{\gamma}_p^\alpha = 0, \tag{11}{}^{\text{d}}$$

where $\dot{\gamma}_p^0$ is the reference shear strain rate, $\Delta G$ is the reference activation energy, $k_B$ is the Boltzmann's constant, $\theta$ is the absolute temperature, $\tilde{s}_l^\alpha$ is the intrinsic lattice resistance, $\langle \bullet \rangle$ is expressed by $\langle \bullet \rangle = \frac{1}{2}(|\bullet| + \bullet)$, $p$ and $q$ are the parameters that control the shape of the energy barrier profile for plastic flow and $\tau_{eff}^\alpha = \tau^\alpha - \tilde{s}^\alpha$ is the effective shear stress, where $\tilde{s}^\alpha$ is the structural resistance due to dislocation evolution and $\tau^\alpha$ is the resolved shear stress projected from the Schmid tensor; i.e., $\tau^\alpha = \mathbf{C}^e \mathbf{T}^* : \mathbf{S}_0^\alpha$. Throughout this work, the resolved shear stress is approximated by $\tau^\alpha \approx \mathbf{T}^* : \mathbf{S}_0^\alpha$ since the elastic distortion is small. Meanwhile, for non-Schmid effects, the effective shear stress is modified to include the non-glide stresses; i.e., $\tau_{eff}^\alpha = \tilde{\tau}^\alpha - \tilde{s}^\alpha$, where $\tilde{\tau}^\alpha = \mathbf{T}^* : \left(\mathbf{S}_0^\alpha + \tilde{\mathbf{S}}_0^\alpha\right)$ is the modified resolved shear stress with the three-term or five-term projection tensors as discussed earlier in **Section 2**. Moreover, the plastic resistances are corrected for a temperature effect as follows,

$$\tilde{s}^\alpha = s^\alpha \frac{\mu}{\mu_0} \text{ and } \tilde{s}_l^\alpha = s_l^\alpha \frac{\mu}{\mu_0}, \tag{12}$$

where $s^\alpha$ and $s_l^\alpha$ are the structural resistance and the intrinsic lattice resistance at 0 K, $\mu_0$ is the anisotropic shear modulus at 0 K and $\mu$ is the anisotropic shear modulus at the current temperature. Moreover, the structural resistance evolves with plastic flow, for which a standard hardening rule was used as follows,

$$\dot{s}^{\alpha} = \sum_{\beta=1}^{N} h_{\alpha\beta} \left| \dot{\gamma}_{p}^{\beta} \right|, \tag{13}$$

where $h_{\alpha\beta}$ is the hardening matrix expressed by $h_{\alpha\beta} = \left(q_l + (1-q_l)\delta_{\alpha\beta}\right)h_{\beta}$. Here, $q_l \in [1, 1.4]$ is the latent hardening parameter and $h_{\beta} = h_0 \left( \dfrac{s_{ss}^{\beta} - s^{\beta}}{s_{ss}^{\beta} - s_0^{\beta}} \right)$ is the self-hardening rate with the parameters of $h_0$, $s_0^{\beta}$ and $s_{ss}^{\beta}$. More generalized and complex hardening laws that account for dislocations interacting throughout the slip systems and their evolutions can be used for these BCC crystals (Cereceda et al., 2016; Madec and Kubin, 2017; Queyreau et al., 2009; Stainier et al., 2002). However, in this work, we have used the classical hardening law for brevity. Further, we restricted our attention to isothermal conditions. Meanwhile, under adiabatic conditions, the rate of heating due to plastic flow is computed by $\rho c \dot{\theta} = \eta \mathbf{C}^e \mathbf{T}^* : \mathbf{L}^p$, where $\rho$ is the density, $c$ is the specific heat, $\mathbf{C}^e \mathbf{T}^* : \mathbf{L}^p = \sum_{\alpha=1}^{N} \tau^{\alpha} \cdot \dot{\gamma}_p^{\alpha}$ is the plastic stress power per unit volume [e], and $\eta \in [0, 1.0]$ is the conversion factor of plastic work to heat. Moreover, the stress should be modified to have thermal expansion; i.e., $\mathbf{T}^* = \mathcal{C}^c \left[ \mathbf{E}^e - (\theta - \theta_0)\mathbf{A} \right]$, where $\mathbf{A}$ is the thermal expansion tensor and $\theta_0$ is the initial temperature.

The constitutive model has been numerically implemented to model the stress-strain behaviors of a representative BCC tantalum. Specifically, we employed an implicit time integration procedure presented by Anand and coworkers (Anand, 2004; Bronkhorst et al., 1992; Kalidindi et al., 1992).

### 3.2. Stress-strain Behavior in Tension and Compression with 3-term and 5-term Formulations

Deformation features of a BCC single crystal are presented to further elucidate non-Schmid effects within the constitutive model that employs the two projection tensors. Specifically, we modeled the stress-strain behaviors of an exemplar BCC single crystal, tantalum (see **Table I** in **Appendix B** for material parameters used in the constitutive model). Moreover, the stress-strain curves were simulated under uniaxial tension and compression along representative crystallographic orientations. Detailed information on symmetric or periodic boundary conditions for a single crystal model can be found in the literature (Cho et al., 2016; Danielsson, 2003; Hansen et al., 2013; Hansen et al., 2010).

**Figure 3** shows simulated stress-strain curves under uniaxial tension and compression with the three-term formulation (strain rate: 0.1 /s and temperature: 200 K). As expected, there was no asymmetry between tension and compression with the Schmid law in all directions. Stress-strain curves for uniaxial stress along [001] are presented in **Figure 3a**, **b** and **c**. As $\omega_1$ (or $\omega_3$) increases, the yield stress in tension was substantially lessened with no corresponding change in compression. In this orientation, the non-Schmid stresses via $\tilde{\mathbf{S}}_0^{1,\alpha}$ and $\tilde{\mathbf{S}}_0^{3,\alpha}$ lead to asymmetry by changing the yield stress in tension. On the other hand, the yield stresses in tension and compression increased simultaneously with an increased $\omega_2$, i.e., the non-Schmid stress via $\tilde{\mathbf{S}}_0^{2,\alpha}$ does not contribute any asymmetry in this orientation but increases the magnitude of the yield stress for both tension and compression. **Figure 3d**, **e** and **f** show the stress-strain curves along [101]. As expected, the non-Schmid stress via $\tilde{\mathbf{S}}_0^{2,\alpha}$ does not lead to any asymmetry. However, the yield stresses for both tension and compression decreased with an increased $\omega_2$, which is contrary to the result for uniaxial stress along [001]. Though the non-Schmid stresses via $\tilde{\mathbf{S}}_0^{1,\alpha}$ and $\tilde{\mathbf{S}}_0^{3,\alpha}$ were again found to be responsible for asymmetry, the asymmetric features are very different from those observed from simulations along [001]. As $\omega_1$ increases, the stress response in compression was significantly lessened with no corresponding change in tension, which is contrary to that observed from simulations along [001]. Moreover, the yield stress in compression increased with an increased $\omega_3$ along [101] while it decreased in tension along [001]. The stress-strain curves for uniaxial stress along [11$\bar{1}$] are presented in **Figure 3g**, **h** and **i**. Overall features are very similar to those along [101] including the tension-compression asymmetry due to $\tilde{\mathbf{S}}_0^{1,\alpha}$ and $\tilde{\mathbf{S}}_0^{3,\alpha}$. However, sensitivities to the weighting factors were found to be much greater than those observed along [101]. In turn, the asymmetric features in tension and compression were found to be more complex along [101] and [11$\bar{1}$] due to the presence of an additional projection plane ($\mathbf{n}'_0$). Moreover, with the simple deviatoric formulation on the non-Schmid projection tensor, the tension-compression asymmetry can be nicely captured without any pressure sensitivity.

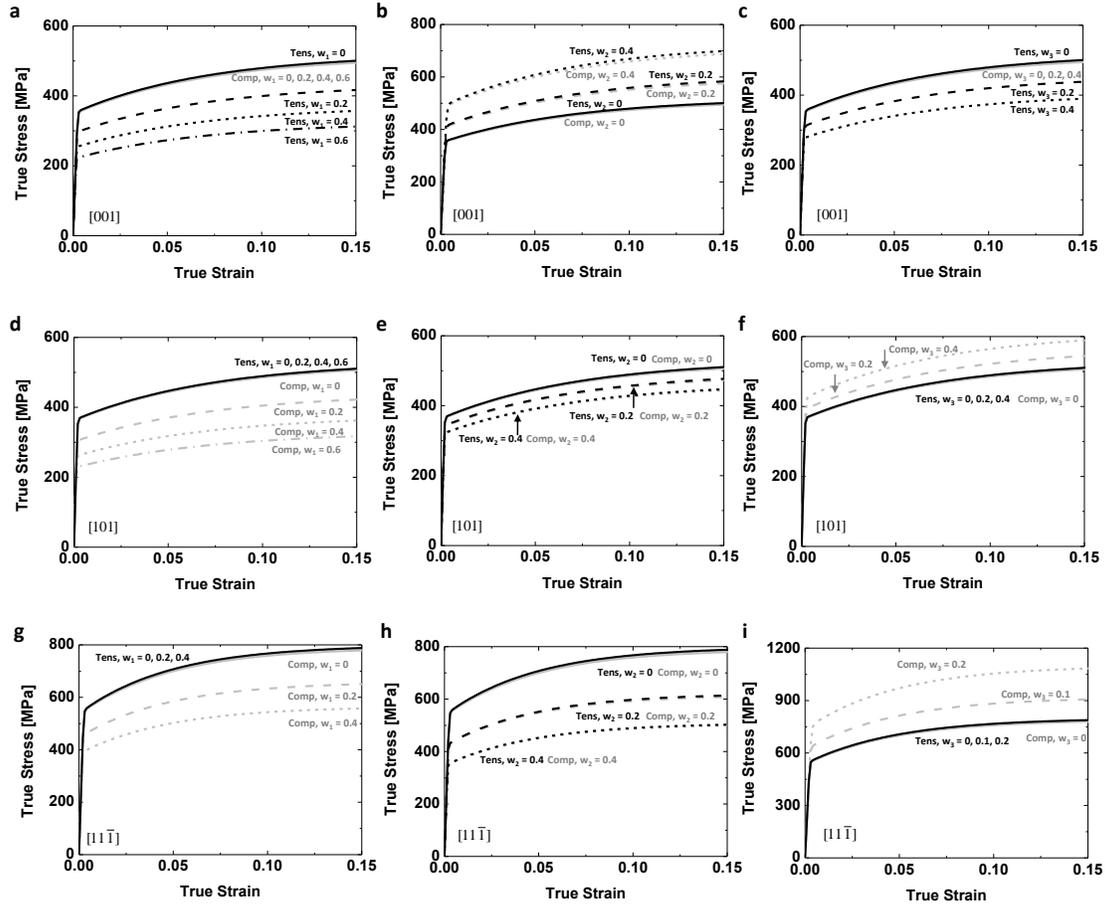

**Figure 3** Stress-strain behaviors of a BCC single crystal under uniaxial tension and compression with varying $\omega_1$, $\omega_2$ and $\omega_3$ (strain rate: 0.1 /s, temperature: 200K). (a) $\omega_1$, (b) $\omega_2$, (c) $\omega_3$ in [001], (d) $\omega_1$, (e) $\omega_2$, (f) $\omega_3$ in [101], (g) $\omega_1$, (h) $\omega_2$, (i) $\omega_3$ in [11$\bar{1}$].

Stress-strain curves of a BCC single crystal were modeled under tension and compression for the five-term formulation in **Figure 4**. As shown in the single slip analysis in **Section 2**, asymmetry between tension and compression were found to be attributed to the non-Schmid stresses via $\tilde{\mathbf{S}}_0^{1,\alpha}$ and $\tilde{\mathbf{S}}_0^{i=3,4,5,\alpha}$. However, as shown in **Figure 4a**, **c**, **d**, **f**, **g** and **i**, a difference between yield stresses with and without non-Schmid effects was almost identical in both compression and tension for these non-Schmid projection tensors, which was very different from that found in the three-term formulation. Moreover, the non-Schmid stress via $\tilde{\mathbf{S}}_0^{2,\alpha}$ did not contribute to asymmetry. For both compression and tension, the yield stress was found to increase or decrease simultaneously due to $\tilde{\mathbf{S}}_0^{2,\alpha}$, which was similar to that in the three-term formulation, i.e., $\tilde{\mathbf{S}}_0^{2,\alpha}\big|_{3-term} = -\tilde{\mathbf{S}}_0^{2,\alpha}\big|_{5-term}$. Notably, the additional "normal" stresses via $\tilde{\mathbf{S}}_0^{i=3,4,5,\alpha}$ resulted in remarkable asymmetry as presented in **Figure 4c**, **f** and **i**, where the constraint, $\omega_3 + \omega_4 + \omega_5 = 0$, was imposed, by which $\sum_{i=3}^{5} \omega_i \tilde{\mathbf{S}}_0^{i,\alpha}$ was still "deviatoric".

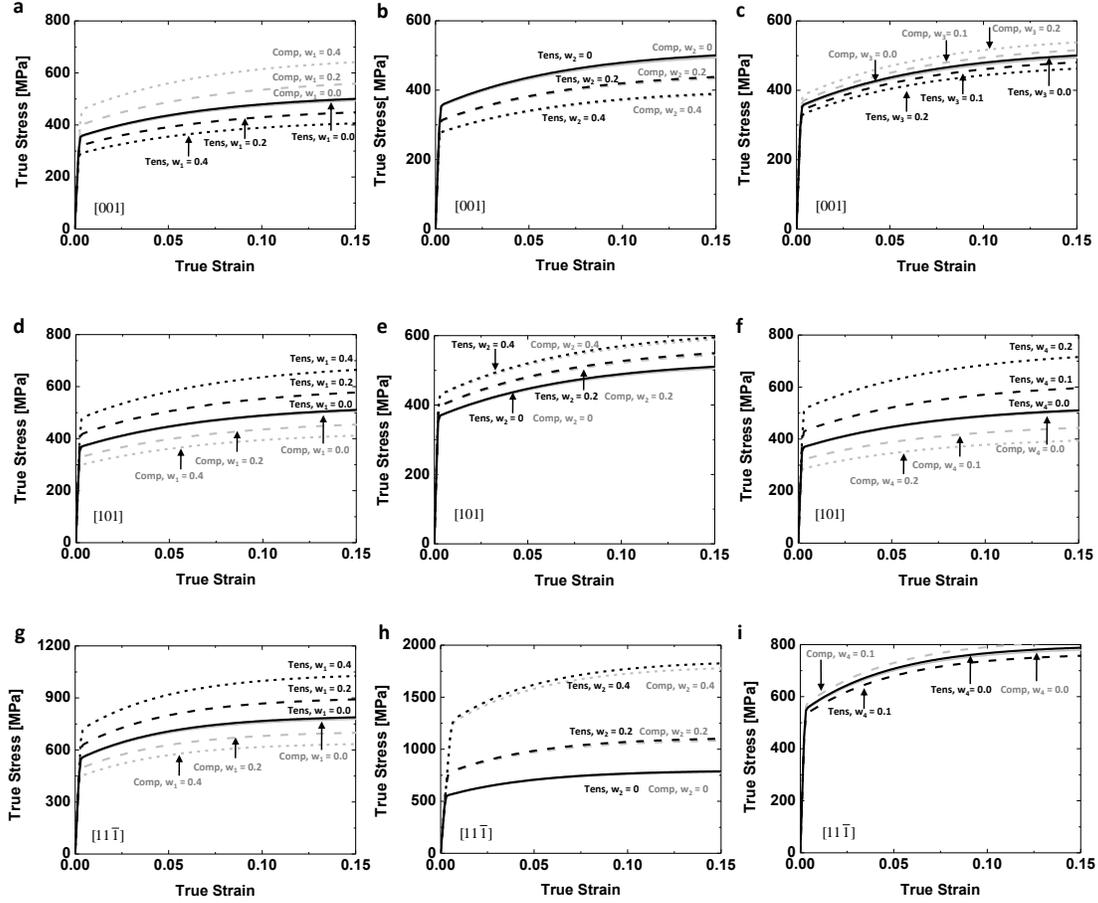

**Figure 4** Stress-strain behaviors of a BCC single crystal under uniaxial tension and compression with varying $\omega_1$, $\omega_2$, $\omega_3$ and $\omega_4$ (strain rate: 0.1 /s, temperature: 200K). (a) $\omega_1$, (b) $\omega_2$, (c) $\omega_3 = -\omega_5$ in [001], (d) $\omega_1$, (e) $\omega_2$, (f) $\omega_4 = -\omega_5$ in [101], (g) $\omega_1$, (h) $\omega_2$, (i) $\omega_4 = -\omega_5$ in $[11\bar{1}]$.

Further, the role of additional normal stresses is addressed with a non-deviatoric projection tensor, $\sum_{i=3}^{5}\omega_i\tilde{\mathbf{S}}_0^{i,\alpha}$, where $\omega_3 + \omega_4 + \omega_5 > 0$. As shown in **Figure 5**, the yield stress in compression increased slightly with an increased $\omega_3 + \omega_4 + \omega_5$ while it decreased in tension, compared to that with the Schmid law, revealing that the non-deviatoric projection tensor, $\sum_{i=3}^{5}\omega_i\tilde{\mathbf{S}}_0^{i,\alpha}$, leads to a pressure-sensitive plastic flow, in particular, when $\omega_3 = \omega_4 = \omega_5$ for $\tilde{\mathbf{S}}_0^{i=3,4,5,\alpha}$. It should be noted that if these weighting factors are not identical, the yield stress is different from that with the identical weighting factors, even though $\omega_3 + \omega_4 + \omega_5$ is the same in both cases; i.e., for pressure-sensitive plastic flows, the constraint, $\omega_3 = \omega_4 = \omega_5$, should be strictly imposed since the hydrostatic effect to plastic flow is coupled with non-Schmid effects from normal stresses without this constraint. Moreover, $\omega_{i=3,4,5}$ should be strictly positive for single crystals with no initial imperfection.

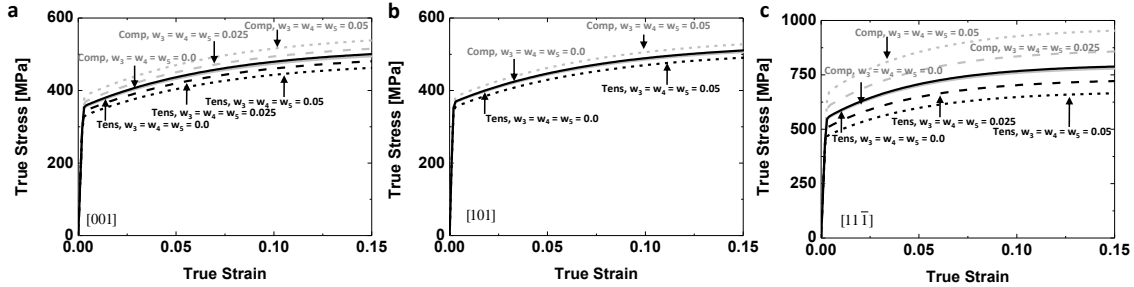

**Figure 5** Stress-strain behaviors of a BCC single crystal with pressure-sensitivity ($\omega_3 = \omega_4 = \omega_5 > 0$). (a) $[001]$, (b) $[101]$ and (c) $[11\bar{1}]$. Here, $\omega_{1,2} = 0$

### 3.3. Stress-strain Behavior of Tantalum: Experiment and Model

Deformations of a representative BCC tantalum are further addressed for both experiment and model at a range of low temperatures. The stress-strain behaviors of tantalum are modeled, in particular, with the three-term formulation for non-Schmid effects along different crystallographic orientations. The experimental data were taken from Sherwood and coworkers (Sherwood et al., 1967), where anomalous tension-compression asymmetry in single crystal tantalum was well described. For non-Schmid effects, the weighting factors were taken to be $\omega_1 = 0.5$, $\omega_2 = 0.2$ and $\omega_3 = 0.1$ at very low temperature (77 K), which are close to those identified in atomistic simulations for tantalum (Alleman et al., 2014). However, they were further tuned to capture the yield stresses at 77 K.[f] **Figure 6a** shows the stress-strain curves of tantalum under tension and compression along $[001]$ in both experiment and model. The single crystal model matched nicely the experimental data for both yield and post-yield hardening behaviors, revealing the ability to capture the significant tension-compression asymmetry in this particular orientation. Here, the stress-strain curves with the Schmid law were presented together (dashed lines for both compression and tension); the symmetric yield stress predicted with the Schmid law was apparently located between the yield stresses for compression and tension with non-Schmid effects. Moreover, **Figure 6b** shows the stress-strain curves under compression along both [001] and [101] in experiment and model. The simulated stress-strain curve along [101] was found to match satisfactorily the experimental data, in which the stresses along [101] were much lower than those along [001] at small to moderate strains.

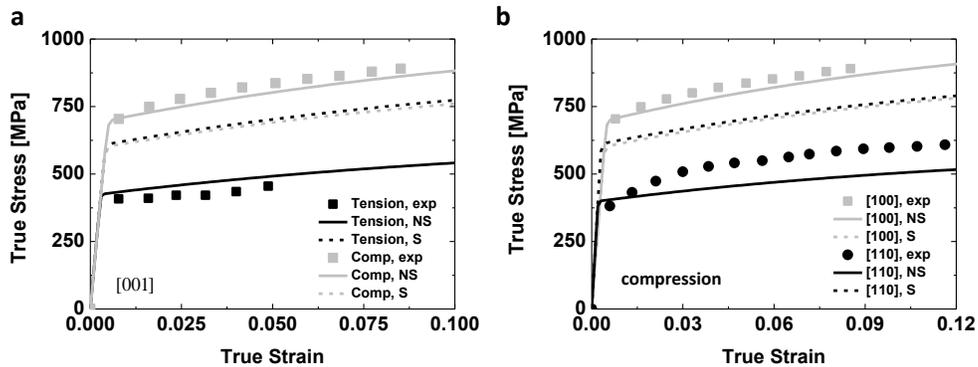

**Figure 6** Stress-strain behavior of tantalum in experiment and model (strain rate: ~ 0.001 /s, temperature: 77 K). (a) Stress-strain curves under tension and compression along [001], (b) Stress-strain curves under compression along [001] and [101]. (NS: model with non-Schmid effect and S: model with Schmid law)

The orientation-dependent yield stresses are further presented for experiment and model in **Figure 7**, in which the yield stresses predicted with non-Schmid effects were shown together with those for the Schmid law. The tension-compression asymmetry was found to be strongly dependent on the crystallographic orientation; i.e. the yield stress in compression was much higher than in tension along [001] while it was lower than in tension along [101]. The constitutive model with non-Schmid effects was, in turn, able to capture reasonably the main features of the orientation-dependent tension-compression asymmetry in the material.

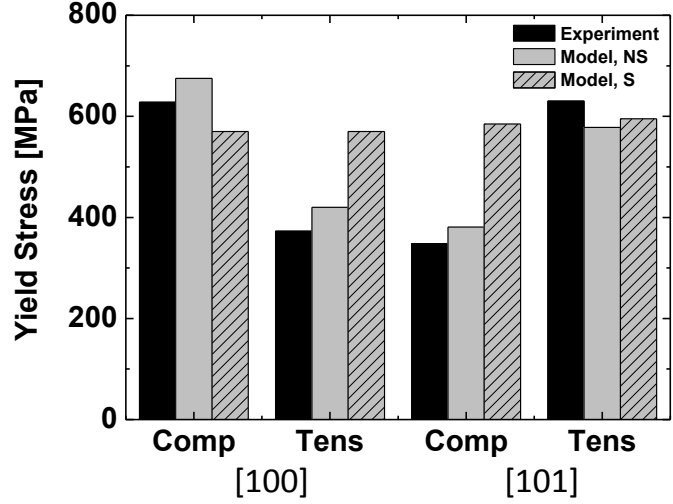

**Figure 7** Orientation-dependent yield properties in tantalum under compression and tension (NS: non-Schmid and S: Schmid)

The non-Schmid effect is substantially dependent on temperature, as well described at continuum and atomistic scales (Cereceda et al., 2016; Chen et al., 2013; Lim et al., 2015; Patra et al., 2014). In many BCC crystals, it has been widely observed that the non-Schmid effect decreases as temperature increases. Moreover, the tension-compression asymmetry in this single crystal tantalum was found to vanish near room temperature. The single crystal constitutive model is hence furthered to capture the temperature-dependent non-Schmid effects, for which the weighting factor is taken to be a function of temperature. Here, we used a simple saturation-type equation for the weighting factor, i.e., $\omega_i = \omega_{i,ss} + (\omega_{i,0} - \omega_{i,ss})\exp(-\theta/\theta_r)$, where $\omega_{i,0}$ is the weighting factor ($i=1,2,3$) at 0 K in the three-term formulation, $\omega_{i,ss}$ is the saturated weighting factor, and $\theta_r$ is the characteristic temperature that controls the rate of decay of non-Schmid effects with temperature. The parameters ($\omega_{i,0}$, $\omega_{i,ss}$ and $\theta_r$) were then identified, especially for a range of low temperatures, $\theta \in [77K, 200K]$, of particular interest in this work. For brevity, the saturated weighting factor ($\omega_{i,ss}$) was taken to be 5% of the weighing factor ($\omega_{i,0}$) at 0 K, which was simply extrapolated from the weighting factor identified at 77 K ($\omega_{1,77K} = 0.5$, $\omega_{2,77K} = 0.2$ and $\omega_{3,77K} = 0.1$). Moreover, the characteristic temperature ($\theta_r$) was estimated using the decaying rate of tension-compression asymmetry in experiment. These parameters are given in **Table I** together with other material parameters used in the constitutive model.

The yield stresses in tension and compression are presented for both experiment and model in **Figure 8**. As evident in the experimental data (**Figure 8a** and **Figure 8c**), the tension-compression asymmetry in this tantalum lessened significantly along both orientations as temperature increased. The saturation-type equations for the weighting factors enable to capture the highly temperature-dependent tension-compression asymmetry, as presented in **Figure 8b** and **Figure 8d**; i.e., the decreased asymmetry with increased temperature was reasonably described in the model.

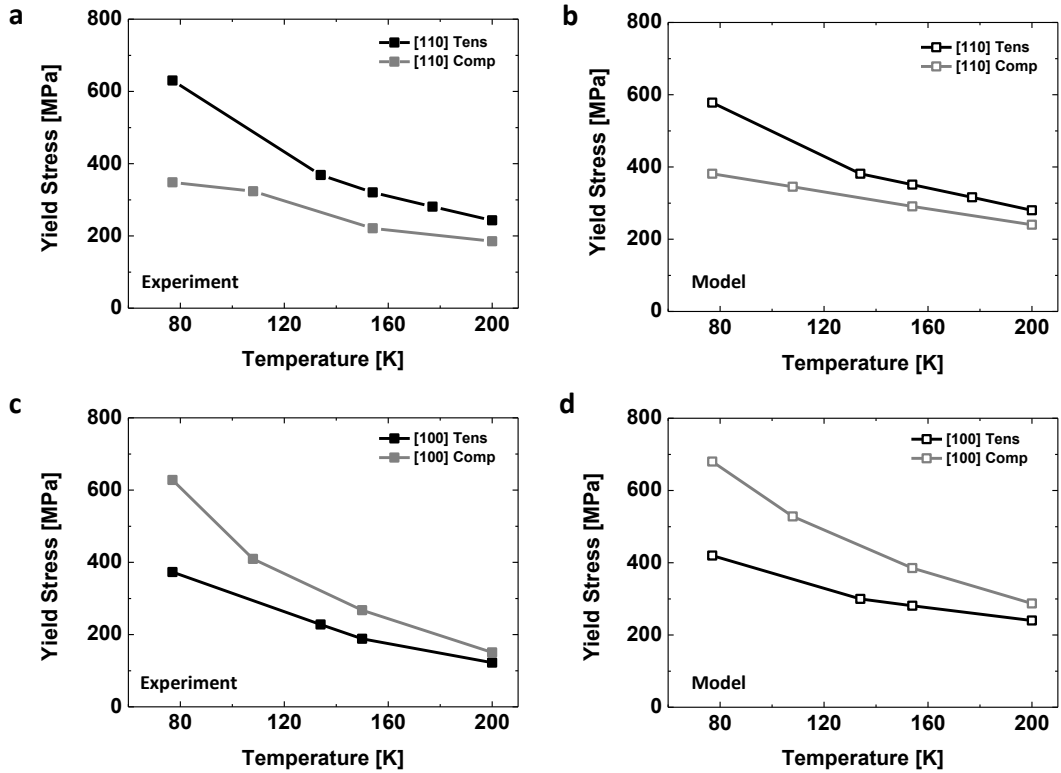

**Figure 8** Temperature-dependent yield stress under tension and compression. (a) Experiment, (b) Model along $[101]$, (c) Experiment, (d) Model along $[001]$

Meanwhile, the tension-compression asymmetry in this tantalum vanished at room temperature, where all of the yield stresses in compression and tension along both [001] and [101] converged near 100 MPa in experiment (Sherwood et al., 1967). However, at room temperature, a yield stress (175 MPa) predicted with the temperature-dependent non-Schmid effect was higher than that in experiment. Further, at > 120 K, the model predictions with or without non-Schmid effects were found to be higher than in experiment for both compression and tension, in particular, along [001], as evident in **Figure 8c** and **Figure 8d**. In this work, we restricted our attention to the single crystal plasticity with non-Schmid effects at relatively low temperatures, for which only $\{110\}\langle 111\rangle$ slip systems have been taken into account. However, in BCC crystals, $\{112\}\langle 111\rangle$ slip systems can be activated at moderate temperature (Seeger, 2001). Therefore, if the $\{112\}\langle 111\rangle$ slip systems are included in the model with appropriate activation temperature for these slips, the model prediction for yield can be lower, due to more slips accumulated on $\{110\}$ or $\{112\}$ planes. Moreover, the weighting factors may be further modified to include plastic strain (or dislocation density) dependence of non-Schmid effects vanishing gradually after the onset of yield.

## 4. Conclusion

In this work, we addressed deformation mechanisms of BCC single crystals that exhibit non-Schmid effects. A comprehensive modeling study on two non-Schmid tensors, widely accepted for representing the non-glide components of an applied stress tensor, showed how each of the non-glide stresses influences abnormal features in plastic flow of these crystals under diverse loading conditions. Moreover, a constitutive model for single crystal plasticity was further specialized to include the non-Schmid effects via the two approaches, by which salient features in tension-compression asymmetry were elucidated. In particular, the three-term formulation was found to be capable of modeling more complex asymmetric behaviors in different crystallographic orientations, compared to the five-term formulation.

Further, stress-strain behaviors of a representative BCC tantalum were simulated using the constitutive modeling framework and compared to experimental data, revealing the abilities of the model to capture the highly orientation- and temperature-dependent non-Schmid effects in the material.

In future work, the non-Schmid effect should be further investigated to span a wide range of strain rates and temperatures in experiment and model for both single- and polycrystalline BCC materials, where the temperature-dependent plastic flow is strongly coupled with the rate-dependence. Moreover, the evolution of dislocations interacting throughout slip systems in the presence of non-Schmid effects should be taken into account, by which the constitutive model can be further improved.

[a] In the studies using atomistic simulations, it is sufficient to examine the critical shear stress on MRSSP with $\varphi \in [-30°, 30°]$ due to the symmetry in these cubic crystals in the zone of a $\langle 111 \rangle$ direction, i.e., the MRSSP is bounded between two $\{112\}$ planes (e.g. $(\bar{1}\bar{1}2)$ and $(\bar{2}11)$ in $\alpha = 2$), which are twinning/anti-twinning planes in BCC crystals.

[b] In these atomistic simulations on a single crystal tungsten, the critical shear stress on MRSSP was found to be symmetric about $\varphi$, in particular, under shear (here, $\tau^*/C_{44} = 0.028$). However, there was an apparent asymmetry under uniaxial compression and tension, which cannot be captured by the Schmid law (see Gröger and coworkers (Gröger et al., 2008a; Gröger et al., 2008b))

[c] In the three-term formulation, the non-Schmid effect does not vanish at $\varphi = 0°$ while it vanishes naturally in the five-term formulation (Equation (3) and (5) with $\omega_{i \neq 1} = 0$). Thus, $\tau^*$ with non-Schmid effects is identified separately from that with the Schmid law, by which the two curves with and without non-Schmid effects meet at $\varphi = 0°$ (**Figure 1c** and **Figure 2a**). However, in the five-term formulation, $\tau^*$ fit at $\varphi = 0°$ can be used for both Schmid and non-Schmid cases (**Figure 1d** and **Figure 2d**).

**d)** Alternatively, the flow rule can be expressed by, as follows,

$$\dot{\gamma}_p^\alpha = \dot{\gamma}_p^0 \exp\left(-\frac{\Delta G}{k_B \theta}\left\langle 1 - \left(\frac{\tau^\alpha}{\tilde{s}^\alpha + \left\langle \tilde{s}_l^\alpha - \mathbf{T}^* : \tilde{\mathbf{S}}_0^\alpha \right\rangle}\right)^p \right\rangle^q \right) \text{ for } \tau^\alpha > 0, \text{ otherwise } \dot{\gamma}_p^\alpha = 0.$$

**e)** When computing the dissipated work density due to plastic flow, the RSS ($\tau^\alpha$) should be used since the non-Schmid stresses do not produce the Peach-Koehler force; i.e., they do not do any work but influence the dislocation core. Therefore, only the RSS does the work through dislocation glide, consistent with the plastic velocity gradient, $\mathbf{L}^p = \sum_{\alpha=1}^{N} \dot{\gamma}_p^\alpha \mathbf{S}_0^\alpha$, that uses the original Schmid tensor.

**f)** The choice for the weighting factors is not unique for this material. Orientation-dependent asymmetry in tantalum with non-Schmid effects can vary considerably (Byron, 1968; Sherwood et al., 1967).

**Acknowledgment**

This research was funded by the LDRD program (LDRD-DR-20170033) at Los Alamos National Laboratory (LANL). LANL is operated by Los Alamos National Security, LLC, for National Nuclear Security Administration of the U.S. Department of Energy.

**Appendix A. Critical Shear Stress under a Combined Loading Condition**

Here, $\tau_{CR}$ (Equation (3)) resolved on MRSSP under $\mathbf{T}|_{\{\mathbf{e}_i\}} = -\sigma \mathbf{e}_1 \otimes \mathbf{e}_1 + \sigma \mathbf{e}_2 \otimes \mathbf{e}_2 + 2\tau \cdot \text{sym}(\mathbf{e}_2 \otimes \mathbf{e}_3)$, is derived. First, the three vectors are expressed in terms of the orthonormal basis, $\{\mathbf{e}_i\}$, in the sip system ($\alpha = 2$),

$$\mathbf{m}_0 = \mathbf{e}_3, \quad \mathbf{n}_0 = -\sin\varphi\, \mathbf{e}_1 + \cos\varphi\, \mathbf{e}_2, \quad \mathbf{n}_0' = -\sin(\varphi+60°)\mathbf{e}_1 + \cos(\varphi+60°)\mathbf{e}_2. \tag{A1}$$

Therefore, the projection tensors can be expressed by,

$$\mathbf{S}_0 = \mathbf{m}_0^\alpha \otimes \mathbf{n}_0^\alpha = \begin{pmatrix} 0 & 0 & 0 \\ 0 & 0 & 0 \\ -\sin\varphi & \cos\varphi & 0 \end{pmatrix}\Bigg|_{\{\mathbf{e}_i\}}, \tag{A2}$$

$$\tilde{\mathbf{S}}_0^1 = \mathbf{m}_0 \otimes \mathbf{n}_0' = \begin{pmatrix} 0 & 0 & 0 \\ 0 & 0 & 0 \\ -\sin(\varphi+60°) & \cos(\varphi+60°) & 0 \end{pmatrix}\Bigg|_{\{\mathbf{e}_i\}}, \tag{A3}$$

$$\tilde{\mathbf{S}}_0^2 = (\mathbf{n}_0 \times \mathbf{m}_0) \otimes \mathbf{n}_0 = \begin{pmatrix} -\sin\varphi\cos\varphi & \cos^2\varphi & 0 \\ \sin^2\varphi & \sin\varphi\cos\varphi & 0 \\ 0 & 0 & 0 \end{pmatrix}\Bigg|_{\{\mathbf{e}_i\}}, \tag{A4}$$

$$\tilde{\mathbf{S}}_0^3 = (\mathbf{n}_0' \times \mathbf{m}_0) \otimes \mathbf{n}_0' = \begin{pmatrix} -\sin(\varphi+60°)\cos(\varphi+60°) & \cos^2(\varphi+60°) & 0 \\ \sin^2(\varphi+60°) & \sin(\varphi+60°)\cos(\varphi+60°) & 0 \\ 0 & 0 & 0 \end{pmatrix}\Bigg|_{\{\mathbf{e}_i\}}. \tag{A5}$$

At the onset of yield (for a rate-independent case), we have,

$$\mathbf{T}:\left(\mathbf{S}_0 + \sum_{i=1}^{3}\omega_i \tilde{\mathbf{S}}_0^i\right) = \tau_{CR}\cos\varphi + \omega_1\tau_{CR}\cos(\varphi+60°) + \omega_2\sigma\sin 2\varphi + \omega_3\sigma\sin 2(\varphi+60°) = \tau^*. \tag{A6}$$

By rearranging Equation (A6), we obtain the critical shear stress resolved on MRSSP expressed by,

$$\frac{\tau_{CR}}{\tau^*} = \frac{1 - \dfrac{\sigma}{\tau^*}\left(\omega_2\sin 2\varphi + \omega_3\sin 2(\varphi+60°)\right)}{\cos\varphi + \omega_1\cos(\varphi+60°)}. \tag{A7}$$

**Appendix B Material Parameters for Single Crystal Constitutive Model**

The material parameters used in the single crystal plasticity model for tantalum are provided in **Table I**. The set of parameters has been well validated, in particular, for deformations of polycrystalline tantalum, as is well described by Anand, Bronkhorst and coworkers (Alleman et al., 2014; Bronkhorst et al., 2007; Kothari and Anand, 1998).

The elastic constants are assumed to be linearly dependent on temperature, as follows,

$$C_{IJ}^c = C_{IJ,0}^c + a_{IJ}\theta, \tag{B1}$$

where $C_{IJ,0}^c$ is the elastic constant at 0 K and $a_{IJ}$ is the slope of a temperature dependence. Moreover, the temperature-dependent structural resistance and intrinsic lattice resistance were expressed by $\tilde{s}^\alpha = s^\alpha \frac{\mu}{\mu_0}$ and $\tilde{s}_l^\alpha = s_l^\alpha \frac{\mu}{\mu_0}$ in Equation (12). Here, the anisotropic shear modulus for temperature correction is expressed by,

$$\mu = \sqrt{C_{44}^c \cdot \left(\frac{C_{11}^c - C_{12}^c}{2}\right)}. \tag{B2}$$

If the anisotropy ratio, $\varsigma \equiv \frac{C_{11}^c - C_{12}^c}{2C_{44}^c} = 1$, it simply leads to an isotropic shear modulus. Moreover, the rate-dependent saturation value of the structural resistance is expressed by,

$$s_{ss}^\alpha = s_{ss,0}^\alpha \left(\frac{\dot{\gamma}_p^\alpha}{\dot{\gamma}_p^0}\right)^\xi \tag{B3}$$

where $s_{ss,0}^\alpha$ is the reference saturation value, $\xi = \frac{k_B \theta}{A}$ is the temperature-dependent exponent, and $A$ is the reference energy.

**Table I** Material parameters for single crystal tantalum

| | | | |
|---|---|---|---|
| $C_{11,0}^{c}$ [GPa] | 268.5 | $s_{0}^{\alpha}$ [MPa] | 50 |
| $C_{12,0}^{c}$ [GPa] | 159.9 | $s_{ss,0}^{\alpha}$ [MPa] | 185 |
| $C_{44,0}^{c}$ [GPa] | 87.1 | $p$ | 0.34 |
| $a_{11}$ [MPa/K] | -24.5 | $q$ | 1.66 |
| $a_{12}$ [MPa/K] | -11.8 | $h_{0}$ [MPa] | 300 |
| $a_{44}$ [MPa/K] | -14.9 | $A$ [J] | $1.0 \times 10^{-18}$ |
| $\dot{\gamma}_{p}^{0}$ [s$^{-1}$] | $1.0 \times 10^{7}$ | $\omega_{i=1,0}$ | 1.022 |
| $\Delta G$ [J] | $2.1 \times 10^{-19}$ | $\omega_{i=2,0}$ | 0.409 |
| $q_{l}$ | 1.4 | $\omega_{i=3,0}$ | 0.204 |
| $s_{l}^{\alpha}$ [MPa] | 530 | $\theta_{r}$ [K] | 100.0 |

**Appendix C. Slip Systems**

Dislocations with a Burgers vector, $\langle 111 \rangle$, on planes of $\{110\}$, $\{112\}$ or $\{123\}$ are mainly responsible for plastic flows in BCC crystals. Therefore, three-fold rotational symmetry through the zone of $\langle 111 \rangle$ in a BCC lattice leads to a unique set of slip systems for each of the slip planes. Further, activation for slips is significantly dependent on temperature and strain rate, as is well documented in the literature (Groves and Kelly, 1963; Hull and Bacon, 2001; Seeger, 2001). The standard slip systems of 12-$\{110\}\langle 111 \rangle$ and 12-$\{112\}\langle 111 \rangle$ have been widely used for modeling BCC metals, as provided in **Table II** and **Table III**. However, we focused on the $\{110\}\langle 111 \rangle$ slip systems with non-Schmid effects since the geometry of $1/2\langle 111 \rangle$ screw dislocations on the $\{110\}$ planes have been better understood in experiments and atomistic simulations at a range of low temperatures, of particular interest in this work. In **Table II**, the slip systems for $\alpha = 13 \sim 24$ are conjugated to those for $\alpha = 1 \sim 12$; i.e., only difference between $\alpha$ and $\alpha+12$ is the sign of the slip direction ($\mathbf{m}_0^{\alpha+12} = -\mathbf{m}_0^{\alpha}$). With the Schmid law, the conjugate pairs are degenerate to the original 12 slip systems. However, for non-Schmid effects, the full set of 24-$\{110\}\langle 111\rangle$ slip systems should be used since the non-Schmid stresses are "asymmetric" between $\alpha$ and $\alpha+12$, as was detailed in **Section 2**

**Table II** Slip systems for $\{110\}\langle 111\rangle$ (Gröger et al., 2008a)

| $\alpha$ | $\mathbf{m}_0^{\alpha}$ | $\mathbf{n}_0^{\alpha}$ | $\mathbf{n}_0'^{\alpha}$ | $\alpha$ | $\mathbf{m}_0^{\alpha}$ | $\mathbf{n}_0^{\alpha}$ | $\mathbf{n}_0'^{\alpha}$ |
|---|---|---|---|---|---|---|---|
| 1 | $[111]$ | $(01\bar{1})$ | $(\bar{1}10)$ | 13 | $[\bar{1}\bar{1}\bar{1}]$ | $(01\bar{1})$ | $(10\bar{1})$ |
| 2 | $[111]$ | $(\bar{1}01)$ | $(0\bar{1}1)$ | 14 | $[\bar{1}\bar{1}\bar{1}]$ | $(\bar{1}01)$ | $(\bar{1}10)$ |
| 3 | $[111]$ | $(1\bar{1}0)$ | $(10\bar{1})$ | 15 | $[\bar{1}\bar{1}\bar{1}]$ | $(1\bar{1}0)$ | $(0\bar{1}1)$ |
| 4 | $[\bar{1}11]$ | $(\bar{1}0\bar{1})$ | $(\bar{1}10)$ | 16 | $[1\bar{1}\bar{1}]$ | $(\bar{1}0\bar{1})$ | $(0\bar{1}1)$ |
| 5 | $[\bar{1}11]$ | $(0\bar{1}1)$ | $(101)$ | 17 | $[1\bar{1}\bar{1}]$ | $(0\bar{1}1)$ | $(\bar{1}\bar{1}0)$ |
| 6 | $[\bar{1}11]$ | $(110)$ | $(01\bar{1})$ | 18 | $[1\bar{1}\bar{1}]$ | $(110)$ | $(101)$ |
| 7 | $[\bar{1}\bar{1}1]$ | $(0\bar{1}\bar{1})$ | $(1\bar{1}0)$ | 19 | $[11\bar{1}]$ | $(0\bar{1}\bar{1})$ | $(\bar{1}0\bar{1})$ |
| 8 | $[\bar{1}\bar{1}1]$ | $(101)$ | $(011)$ | 20 | $[11\bar{1}]$ | $(101)$ | $(1\bar{1}0)$ |
| 9 | $[\bar{1}\bar{1}1]$ | $(\bar{1}10)$ | $(\bar{1}0\bar{1})$ | 21 | $[11\bar{1}]$ | $(\bar{1}10)$ | $(011)$ |
| 10 | $[1\bar{1}1]$ | $(10\bar{1})$ | $(110)$ | 22 | $[\bar{1}1\bar{1}]$ | $(10\bar{1})$ | $(0\bar{1}\bar{1})$ |
| 11 | $[1\bar{1}1]$ | $(011)$ | $(\bar{1}01)$ | 23 | $[\bar{1}1\bar{1}]$ | $(011)$ | $(110)$ |
| 12 | $[1\bar{1}1]$ | $(\bar{1}\bar{1}0)$ | $(0\bar{1}\bar{1})$ | 24 | $[\bar{1}1\bar{1}]$ | $(\bar{1}\bar{1}0)$ | $(\bar{1}01)$ |

**Table III** Slip systems for $\{112\}\langle 111\rangle$

| $\beta$ | $\mathbf{m}_0^{\beta}$ | $\mathbf{n}_0^{\beta}$ | $\beta$ | $\mathbf{m}_0^{\beta}$ | $\mathbf{n}_0^{\beta}$ |
|---|---|---|---|---|---|
| 13 | $[111]$ | $(11\bar{2})$ | 19 | $[\bar{1}\bar{1}1]$ | $(112)$ |
| 14 | $[111]$ | $(1\bar{2}1)$ | 20 | $[\bar{1}\bar{1}1]$ | $(\bar{1}21)$ |
| 15 | $[111]$ | $(\bar{2}11)$ | 21 | $[\bar{1}\bar{1}1]$ | $(2\bar{1}1)$ |
| 16 | $[\bar{1}11]$ | $(1\bar{1}2)$ | 22 | $[1\bar{1}1]$ | $(\bar{1}12)$ |
| 17 | $[\bar{1}11]$ | $(12\bar{1})$ | 23 | $[1\bar{1}1]$ | $(121)$ |
| 18 | $[\bar{1}11]$ | $(211)$ | 24 | $[1\bar{1}1]$ | $(21\bar{1})$ |

**Reference**

Alleman, C., Ghosh, S., Luscher, D., Bronkhorst, C.A., 2014. Evaluating the effects of loading parameters on single-crystal slip in tantalum using molecular mechanics. Philosophical Magazine 94, 92-116.
Anand, L., 2004. Single-crystal elasto-viscoplasticity: application to texture evolution in polycrystalline metals at large strains. Computer Methods in Applied Mechanics and Engineering 193, 5359-5383.
Anand, L., Kothari, M., 1996. A computational procedure for rate-independent crystal plasticity. Journal of the Mechanics and Physics of Solids 44, 525-558.
Asaro, R.J., 1983a. Crystal plasticity. Journal of Applied Mechanics 50, 921-934.
Asaro, R.J., 1983b. Micromechanics of crystals and polycrystals. Advances in Applied Mechanics 23, 1-115.
Asaro, R.J., Needleman, A., 1985. Overview no. 42 Texture development and strain hardening in rate dependent polycrystals. Acta Metallurgica 33, 923-953.
Asaro, R.J., Rice, J., 1977. Strain localization in ductile single crystals. Journal of the Mechanics and Physics of Solids 25, 309-338.
Bronkhorst, C., Hansen, B., Cerreta, E., Bingert, J., 2007. Modeling the microstructural evolution of metallic polycrystalline materials under localization conditions. Journal of the Mechanics and Physics of Solids 55, 2351-2383.
Bronkhorst, C., Kalidindi, S., Anand, L., 1992. Polycrystalline plasticity and the evolution of crystallographic texture in FCC metals. Philosophical Transactions of the Royal Society of London A 341, 443-477.
Busso, E.P., 1990. Cyclic deformation of monocrystalline nickel aluminide and high temperature coatings. Massachusetts Institute of Technology.
Byron, J., 1968. Plastic deformation of tantalum single crystals: II. The orientation dependence of yield. Journal of the Less Common Metals 14, 201-210.
Carroll, J., Clark, B., Buchheit, T., Boyce, B., Weinberger, C., 2013. An experimental statistical analysis of stress projection factors in BCC tantalum. Materials Science and Engineering: A 581, 108-118.
Cereceda, D., Diehl, M., Roters, F., Raabe, D., Perlado, J.M., Marian, J., 2016. Unraveling the temperature dependence of the yield strength in single-crystal tungsten using atomistically-informed crystal plasticity calculations. International Journal of Plasticity 78, 242-265.
Chen, Z., Mrovec, M., Gumbsch, P., 2013. Atomistic aspects of screw dislocation behavior in α-iron and the derivation of microscopic yield criterion. Modelling and Simulation in Materials Science and Engineering 21, 055023.
Cho, H., Weaver, J.C., Pöselt, E., Boyce, M.C., Rutledge, G.C., 2016. Engineering the Mechanics of Heterogeneous Soft Crystals. Advanced Functional Materials 26, 6938-6949.
Cuitino, A.M., Ortiz, M., 1993. Computational modelling of single crystals. Modelling and Simulation in Materials Science and Engineering 1, 225.
Danielsson, M., 2003. Micromechanics, macromechanics and constitutive modeling of the elasto-viscoplastic deformation of rubber-toughened glassy polymers, Mechanical Engineering. Massachusetts Institute of Technology, Cambridge.
Dao, M., Asaro, R., 1993. Non-Schmid effects and localized plastic flow in intermetallic alloys. Materials Science and Engineering: A 170, 143-160.
Dezerald, L., Ventelon, L., Clouet, E., Denoual, C., Rodney, D., Willaime, F., 2014. Ab initio modeling of the two-dimensional energy landscape of screw dislocations in bcc transition metals. Physical Review B 89, 024104.
Duesbery, M., Vitek, V., 1998a. Plastic anisotropy in bcc transition metals. Acta Materialia 46, 1481-1492.
Duesbery, M., Vitek, V., Bowen, D., 1973. The effect of shear stress on the screw dislocation core structure in body-centred cubic lattices, Proceedings of the Royal Society of London A: Mathematical, Physical and Engineering Sciences. The Royal Society, pp. 85-111.

Duesbery, M.a.-S., Vitek, V., 1998b. Plastic anisotropy in bcc transition metals. Acta Materialia 46, 1481-1492.
Ferriss, D., Rose, R., Wulff, J., 1962. Deformation of tantalum single crystals. Transactions of the Metallurgical Society of AIME 224, 975-&.
Frederiksen, S.L., Jacobsen, K.W., 2003. Density functional theory studies of screw dislocation core structures in bcc metals. Philosophical Magazine 83, 365-375.
Gröger, R., Bailey, A., Vitek, V., 2008a. Multiscale modeling of plastic deformation of molybdenum and tungsten: I. Atomistic studies of the core structure and glide of 1/2< 111> screw dislocations at 0K. Acta Materialia 56, 5401-5411.
Gröger, R., Racherla, V., Bassani, J., Vitek, V., 2008b. Multiscale modeling of plastic deformation of molybdenum and tungsten: II. Yield criterion for single crystals based on atomistic studies of glide of 1/2< 111> screw dislocations. Acta Materialia 56, 5412-5425.
Groves, G., Kelly, A., 1963. Independent slip systems in crystals. Philosophical Magazine 8, 877-887.
Gurtin, M.E., 2000. On the plasticity of single crystals: free energy, microforces, plastic-strain gradients. Journal of the Mechanics and Physics of Solids 48, 989-1036.
Gurtin, M.E., Fried, E., Anand, L., 2010. The mechanics and thermodynamics of continua. Cambridge University Press.
Hansen, B., Beyerlein, I., Bronkhorst, C., Cerreta, E., Dennis-Koller, D., 2013. A dislocation-based multi-rate single crystal plasticity model. International Journal of Plasticity 44, 129-146.
Hansen, B., Bronkhorst, C., Ortiz, M., 2010. Dislocation subgrain structures and modeling the plastic hardening of metallic single crystals. Modelling and Simulation in Materials Science and Engineering 18, 055001.
Hull, D., Bacon, D.J., 2001. Introduction to dislocations. Butterworth-Heinemann.
Irwin, G., Guiu, F., Pratt, P., 1974. The influence of orientation on slip and strain hardening of molybdenum single crystals. Physica Status Solidi (A) 22, 685-698.
Ismail-Beigi, S., Arias, T., 2000. Ab initio study of screw dislocations in Mo and Ta: a new picture of plasticity in bcc transition metals. Physical Review Letters 84, 1499.
Ito, K., Vitek, V., 2001. Atomistic study of non-Schmid effects in the plastic yielding of bcc metals. Philosophical Magazine A 81, 1387-1407.
Kalidindi, S.R., Bronkhorst, C.A., Anand, L., 1992. Crystallographic texture evolution in bulk deformation processing of FCC metals. Journal of the Mechanics and Physics of Solids 40, 537-569.
Kim, J.-Y., Jang, D., Greer, J.R., 2010. Tensile and compressive behavior of tungsten, molybdenum, tantalum and niobium at the nanoscale. Acta Materialia 58, 2355-2363.
Knezevic, M., Beyerlein, I.J., Lovato, M.L., Tomé, C.N., Richards, A.W., McCabe, R.J., 2014. A strain-rate and temperature dependent constitutive model for BCC metals incorporating non-Schmid effects: application to tantalum–tungsten alloys. International Journal of Plasticity 62, 93-104.
Kocks, U., 1976. Laws for work-hardening and low-temperature creep. Journal of Engineering Materials and Technology 98, 76-85.
Koester, A., Ma, A., Hartmaier, A., 2012. Atomistically informed crystal plasticity model for body-centered cubic iron. Acta materialia 60, 3894-3901.
Kothari, M., Anand, L., 1998. Elasto-viscoplastic constitutive equations for polycrystalline metals: application to tantalum. Journal of the Mechanics and Physics of Solids 46, 5169-6783.
Li, J., Wang, C.-Z., Chang, J.-P., Cai, W., Bulatov, V.V., Ho, K.-M., Yip, S., 2004. Core energy and Peierls stress of a screw dislocation in bcc molybdenum: A periodic-cell tight-binding study. Physical Review B 70, 104113.
Lim, H., Hale, L., Zimmerman, J., Battaile, C., Weinberger, C., 2015. A multi-scale model of dislocation plasticity in α-Fe: incorporating temperature, strain rate and non-schmid effects. International Journal of Plasticity 73, 100-118.
Lim, H., Weinberger, C.R., Battaile, C.C., Buchheit, T.E., 2013. Application of generalized non-Schmid yield law to low-temperature plasticity in bcc transition metals. Modelling and Simulation in Materials Science and Engineering 21, 045015.

Madec, R., Kubin, L.P., 2017. Dislocation strengthening in FCC metals and in BCC metals at high temperatures. Acta Materialia 126, 166-173.

Miehe, C., 1996. Exponential map algorithm for stress updates in anisotropic multiplicative elastoplasticity for single crystals. International Journal for Numerical Methods in Engineering 39, 3367-3390.

Mrovec, M., Nguyen-Manh, D., Pettifor, D.G., Vitek, V., 2004. Bond-order potential for molybdenum: Application to dislocation behavior. Physical Review B 69, 094115.

Nabarro, F.R., Duesbery, M.S., 2002. Dislocations in solids. Elsevier.

Patra, A., Zhu, T., McDowell, D.L., 2014. Constitutive equations for modeling non-Schmid effects in single crystal bcc-Fe at low and ambient temperatures. International Journal of Plasticity 59, 1-14.

Peirce, D., Asaro, R.J., Needleman, A., 1983. Material rate dependence and localized deformation in crystalline solids. Acta Metallurgica 31, 1951-1976.

Qin, Q., Bassani, J.L., 1992a. Non-associated plastic flow in single crystals. Journal of the Mechanics and Physics of Solids 40, 835-862.

Qin, Q., Bassani, J.L., 1992b. Non-Schmid yield behavior in single crystals. Journal of the Mechanics and Physics of Solids 40, 813-833.

Queyreau, S., Monnet, G., Devincre, B., 2009. Slip systems interactions in α-iron determined by dislocation dynamics simulations. International Journal of Plasticity 25, 361-377.

Racherla, V., Bassani, J., 2006. Strain burst phenomena in the necking of a sheet that deforms by non-associated plastic flow. Modelling and Simulation in Materials Science and Engineering 15, S297.

Ravelo, R., Germann, T.C., Guerrero, O., An, Q., Holian, B.L., 2013. Shock-induced plasticity in tantalum single crystals: Interatomic potentials and large-scale molecular-dynamics simulations. Physical Review B 88, 134101.

Savage, D.J., Beyerlein, I.J., Knezevic, M., 2017. Coupled texture and non-Schmid effects on yield surfaces of body-centered cubic polycrystals predicted by a crystal plasticity finite element approach. International Journal of Solids and Structures.

Seeger, A., 2001. Why anomalous slip in body-centred cubic metals? Materials Science and Engineering: A 319, 254-260.

Sherwood, P., Guiu, F., Kim, H.-C., Pratt, P.L., 1967. Plastic anisotropy of tantalum, niobium, and molybdenum. Canadian Journal of Physics 45, 1075-1089.

Stainier, L., Cuitiño, A.M., Ortiz, M., 2002. A micromechanical model of hardening, rate sensitivity and thermal softening in bcc single crystals. Journal of the Mechanics and Physics of Solids 50, 1511-1545.

Staroselsky, A., Anand, L., 2003. A constitutive model for hcp materials deforming by slip and twinning: application to magnesium alloy AZ31B. International Journal of Plasticity 19, 1843-1864.

Thamburaja, P., Anand, L., 2001. Polycrystalline shape-memory materials: effect of crystallographic texture. Journal of the Mechanics and Physics of Solids 49, 709-737.

Vitek, V., 1974. Theory of the core structures of dislocations in BCC metals. Cryst. Lattice Defects 5, 1-34.

Vitek, V., 2004. Core structure of screw dislocations in body-centred cubic metals: relation to symmetry and interatomic bonding. Philosophical Magazine 84, 415-428.

Vitek, V., Mrovec, M., Bassani, J., 2004a. Influence of non-glide stresses on plastic flow: from atomistic to continuum modeling. Materials Science and Engineering: A 365, 31-37.

Vitek, V., Mrovec, M., Gröger, R., Bassani, J., Racherla, V., Yin, L., 2004b. Effects of non-glide stresses on the plastic flow of single and polycrystals of molybdenum. Materials Science and Engineering: A 387, 138-142.

Vitek, V., Perrin, R., Bowen, D., 1970. The core structure of ½ (111) screw dislocations in bcc crystals. Philosophical Magazine 21, 1049-1073.

Woodward, C., Rao, S., 2002. Flexible ab initio boundary conditions: simulating isolated dislocations in bcc Mo and Ta. Physical Review Letters 88, 216402.

Xu, W., Moriarty, J.A., 1996. Atomistic simulation of ideal shear strength, point defects, and screw dislocations in bcc transition metals: Mo as a prototype. Physical Review B 54, 6941.